\documentclass[conference,onecolumn,10pt]{IEEEtran}
\IEEEoverridecommandlockouts

\usepackage{slashbox}
\usepackage{multirow}
\usepackage{amsfonts,amsmath,amssymb}
\usepackage{graphicx}
\usepackage{url}
\usepackage{color}

\newcommand{\be}{\begin{eqnarray}}
\newcommand{\ee}{\end{eqnarray}}

\def\tK{\tilde{K}}

\def\ie{{\em i.e.},~}
\def\eg{{\em e.g.},~}
\def\cf{{\em cf.}~}

\def\E{{\sf E}}

\def \ux {{\underline{x}}}

\newcommand{\ben}{\begin{enumerate}}
\newcommand{\een}{\end{enumerate}}
\newcommand{\beq}{\begin{equation}}
\newcommand{\eeq}{\end{equation}}
\newcommand{\beqa}{\begin{eqnarray*}}
\newcommand{\eeqa}{\end{eqnarray*}}
\newcommand{\bit}{\begin{itemize}}
\newcommand{\eit}{\end{itemize}}
\newcommand{\bt}{\begin{tabular}{c}}
\newcommand{\btt}{\begin{tabular}}
\newcommand{\et}{\end{tabular}}

\newtheorem{proposition}{Proposition}
\newtheorem{definition}{Definition}
\newtheorem{corollary}{Corollary}

\begin{document}

\title{Scheduling Distributed Resources\\ in Heterogeneous Private 
Clouds\thanks{This research was supported in part by NSF CNS 1526133, NSF CNS 1717571 and a Cisco Systems URP gift. 
}}

\author{
\begin{tabular}{cc}
G. Kesidis, Y. Shan, A. Jain, B. Urgaonkar & J. Khamse-Ashari and I. Lambadaris\\
School of EECS, PSU, State College, PA & SCE Dept, Carleton Univ., Ottawa, Canada\\
\{gik2,yxs182,axj182,buu1\}@psu.edu  &
\{jalalkhamseashari,ioannis\}@sce.carleton.ca
\end{tabular}
}

\maketitle

\begin{abstract}
We first consider the static problem of allocating resources to 
(\ie scheduling) multiple distributed application frameworks,
possibly with different priorities and server preferences,
in a private cloud with heterogeneous servers. 
Several fair scheduling mechanisms have been proposed for this purpose. 
We extend prior results on
max-min and proportional fair scheduling to this 
constrained multiresource and multiserver case
for generic fair scheduling criteria.
The task efficiencies (a metric related to proportional
fairness) of max-min fair allocations  found 
by progressive filling
are  compared by illustrative examples.
They show that 
``server specific" fairness criteria and
those that are based on residual (unreserved)
resources are more efficient.

\end{abstract}

\section{Introduction and background}

We consider a  cloud provider that needs to run multiple software
applications on its IT infrastructure. These applications may be distributed and
are also called frameworks or workloads in the literature.
The cloud provider's infrastructure consists of multiple servers connected by a network.
A server may be a physical machine or 
virtual machine (\eg an instance or a container). A server is also referred to as a worker or a slave in some 
popular resource management solutions. Each framework desires multiple IT resources (CPU, memory, network bandwidth, {\em etc.}) for each
of its ``tasks.'' A task is a framework-specific basic unit of work that must be placed within a single server at a given time
(\eg it is useless for a  task to be allocated CPU from one server and memory from another).
The provider's challenge then is to determine who should get how many resources from
which servers. Our interest is in a {\em private} cloud setting wherein notions of fairness have often been used
as the basis for this resource allocation problem. 
In a public setting, on the other hand, the provider's goal is typically to maximize its profit.  

What are meaningful notions of fairness for such {\it multi-resource and multi-server} settings?
This question has received much attention in the recent past. 
Proposed fair schedulers include Dominant Resource
Fairness (DRF) \cite{DRF} extended to multiple servers\footnote{DRF
was originally defined for a single
server in \cite{DRF}. The multiple-server version, called DRFH in
\cite{BLi15,Friedman14}, is also commonly  called just DRF as done in Apache Mesos~\cite{Mesos}
and as we do herein also.}, Task Share Fairness (TSF) \cite{BLi16b}, 
Per Server Dominant Share Fairness (PS-DSF)
\cite{PS-DSF-arxiv,PS-DSF-arxiv2,Jalal18},
among others, \eg \cite{HUG}.
DRF is resource based, whereas TSF and
``containerized" DRF \cite{Friedman14} are
task based\footnote{Containerized DRF has a 
``sharing-incentive" property not possessed by DRF,
and TSF possesses
 ``strategy-proofness" and ``envy-freeness" properties 
which are not possessed by  containerized DRF \cite{BLi16b}.
Unlike DRF and TSF, PS-DSF 
is not necessarily Pareto optimal but is ``bottleneck" fair.
These properties are not addressed herein}.
In the following, we additionally 
consider variants 
of these
schedulers that employ current residual (unreserved)
capacities of the servers in the fairness criteria 
(somewhat similar to ``best fit" variants \cite{BLi15}).

{\bf Background on existing approaches and  their assumptions:}  Typically 
static problem formulations are considered under a variety of simplifying assumptions 
on framework behavior that we discuss below:
\begin{itemize}
\item
  It is assumed that frameworks congest all the available servers. That is, it is assumed that there is sufficient work
  to completely occupy at least one resource in every server.
\item
  It is also assumed that the frameworks' required resources (presumably to achieve certain
performance needs) are well characterized, \eg
\cite{Doyle:2003:MRP:1251460.1251465,Levy2003,Menasce:2003:WSS:1050672.1050719,Chandra:2003:DRA:781027.781067,N.Bennani:2005:RAA:1078027.1078472,159,Cohen:2004:CID:1251254.1251270,Abdelzaher:2002:PGW:506156.506162,Lu:2001:FCA:882481.883781}. 
\item
  Frameworks are assumed to have linearly elastic resource demands in the following sense.  
   Each task has a known requirement 
  $d_{n,r}$ for 
   the resource type $r$. Therefore, if $x_{n,i}$ were the number of tasks of framework $n$ placed on server $i$, the framework
   would consume $x_{n,i}d_{n,r}$ amount of resource $r$ on server $i$. 
\item   $x_{n,i}$ may take on non-negative real values rather than 
being restricted to be non-negative integer valued\footnote{With $x$ integer valued, 
such problems belong to the class of combinatorial-optimization
multidimensional bin-packing problems,
\eg \cite{CK04,CKPT16,Cohen17}, 
which are NP-hard. They have been extensively studied, including relaxations to
simplified problems that yield approximately optimal
solutions, \eg by Integer Linear Programs solved by
iterated/online means.}. 
\item
  Finally, frameworks may have different service priorities and server preference constraints (as in \eg service-quality
  constraints  \cite{McKeown13} or cache-affinity constraints), see also \cite{BLi16b}.
\end{itemize}

Note that in some settings, a goal is to minimize the
number of servers to accommodate workloads with finite needs,
again as in
multidimensional bin-packing problems
\cite{CK04,CKPT16,Cohen17}.
Such problem formulations 
are typically motivated by the desire to economize on
energy. However, frequently cycling power to (booting up) servers  may 
result in software errors 
and there are energy spikes associated with boot-up resulting
in increased electricity costs \cite{duke}. 
We are not interested in such settings herein.

Typically in 
existing papers,  max-min fairness with respect to a proposed 
fairness criteria is  specified
assuming the aforementioned congested regime under the following
(linear) capacity constraints:
\be
\forall i,r,~~
\sum_n  x_{n,i}d_{n,r} & \leq & c_{i,r} ,
\label{capacity-constraints}
\ee
where $c_{i,r}$ is the amount of available resource $r$ in server $i$
for the instances under consideration\footnote{Note that if 
$x_{n,i}=0$ then workload type $n$ is not assigned to server $i$.}.
Additionally, there may be  placement constraints,
$\delta_{n,i}\in\{0,1\}$, whereby 
$x_{n,i}>0$ $\Rightarrow$ $\delta_{n,i}=1$.
Max-min fair allocation may be expressed as
the solution of a constrained centralized optimization problem.
Alternatively, max-min fairness with respect to the
proposed  fairness criteria may be
 approximated by a greedy, iterative
``progressive filling" allocation.
The latter  approach is often preferred  because of the 
benefits this offers for online
implementations. Moreover, progressive filling arguments
can be used to 
establish other potentially desirable fairness properties of schedulers
defined for private clouds\footnote{Again, Pareto optimality, sharing incentive, strategy 
proofness, bottleneck fairness, and envy freeness  \cite{DRF} - properties that are
not addressed herein.}. 

Instead of max-min fairness,
the cloud may admit and place instances 
so as to maximize, \eg
total weighted tasking objective,
\be\label{revenue}
\sum_n \phi_n \sum_i x_{n,i}
\ee
subject to
(\ref{capacity-constraints}), where
$\phi_n>0$ is the priority of application framework $n$.
In this paper, we relate this task efficiency
objective to ``proportional" fairness.

In Sections \ref{sec:mmf-theorem} and \ref{sec:PF},
for generic fairness criteria,
we generalize to multiple resources
the static optimization problems
of 
\eg \cite{BG92,Mo00,Ashari16b} 
whose solutions
correspond to max-min fairness  and proportional fairness, 
respectively.
In Section \ref{sec:progfill}, a simple, greedy, iterative method
intended to achieve max-min fairness called progressive filling is described.
Progressive filling is important for online implementation.
In Section \ref{sec:evalobj}, the performance evaluation objectives of the
following two sections are discussed: task efficiency (related to proportional
fairness) and overall execution time.
In Section \ref{sec:numer},
illustrative numerical  examples are used
to compare the task efficiencies of different schedulers,
including variants using residual/unreserved server 
resource capacities specified herein. 
In \cite{Spark-Mesos-arxiv}, we give the results of
an online experimental study using our implementations of different schedulers
on Spark and Mesos 
\cite{mesos-code,spark-code}
for benchmark workloads considering an execution-time
performance metric.
The paper concludes with a summary in Section \ref{sec:summary}
and a brief discussion of future work (regarding scheduling in
public clouds).

Our mathematical notation is given in Table \ref{symbols}.
\begin{table}[h!]
\centering
	\begin{tabular}{|c|c|}
		\hline
		Symbol & Definition \\
		\hline
		$$
		$i$ & server index \\
		$n$ & user/framework index \\
		$r$ & resource type index \\
		$\rho$ & index of the dominant resource \\ 
		$\phi_n$ & weight/priority of user $n$ \\
		$x_{n,i}$ & the number of tasks or workload intensity \\
		$d_{n, r}$ & per-task resource requirement  \\
		$c_{i, r}$ & the total available resource amounts  \\
		$B_{n,i, r}$ & $=d_{n,r}/c_{i,r}$\\
		$\delta_{n,i}$ & server preference indicator \\
		$N_i$ & the set of users that can run on server $i$ \\
		$R_i$ & fully booked resources of server $i$ under $x$\\
		$U_n,K_n,M_n$ & allocation-fairness scores  \\
		\hline
	\end{tabular}
	\caption{Mathematical notation.}\label{symbols}
\end{table}

\section{Max-Min Fairness} \label{sec:mmf-theorem}

To generalize 
previous results on max-min fairness
(\eg \cite{BG92,DRF,Ashari16b})
to multiple resource types on multiple servers,
consider the following general-purpose fairness criterion
for framework $n$,
\be\label{generic-criterion}
U_n & = & \frac{1}{\phi_n}\sum_i u_{n,i}x_{n,i},
\ee
for scalars $u_{n,i}> 0$ and priorities $\phi_n>0$
(specific examples of fairness criteria are given below).
In addition, consider the service-preference sets  
\be\label{x-sum-def}
N_i =\{ n ~|~\delta_{n,i}=1\} & \mbox{where} & 
x_{n,i}>0~\Rightarrow~\delta_{n,i}>0.
\ee
Relaxing the allocations $\{x_{n,i}\}$ to be real valued,
consider strictly concave and increasing $g$ with $g(0)=0$,  and
define the optimization problem
\be\label{Omega2}
\max_{x}\sum_n \phi_n g(U_n)  
\ee
such that (here restating (\ref{capacity-constraints}))
\be
\forall i,r, ~\sum_{n\in N_i} x_{n,i}B_{n,i,r}  \leq  1
& \mbox{and}&  \forall n,i ~x_{n,i}  \geq  0, 
\label{no-overbook2}
\ee
where
\be
B_{n,i,r}  & := & \frac{d_{n,r}}{c_{i,r}}. \label{B-def}
\ee
Note that the objective is continuous and strictly concave and the 
domain given by (\ref{no-overbook2}) (equivalently
(\ref{capacity-constraints})) is compact. 
So,   simply by Weierstrass's Extreme Value Theorem,
there exists a unique maximum.

Regarding fully booked resources in server $i$ under
allocations $x=\{x_{n,i}\}$, also let
\beqa
R_i & := & \{(x,r) ~|~\sum_{n\in N_i} x_{n,i}B_{n,i,r}=1 \}.
\eeqa
For the following definition, assume that
$\forall n,i,r,~ B_{n,i,r}>0$.

\begin{definition}
A feasible allocation $\{x_{n,i}\}$ satisfying  
(\ref{no-overbook2})  is said to be
$U$-Max-Min Fair (MMF) if:
\beqa
U_\ell >U_m,~
x_{m,i}>0,~\&~\exists r~\mbox{s.t.}
\sum_{n\in N_i}x_{n,i}B_{n,i,r}=1
\eeqa
implies that $x_{\ell,i}=0$.
\end{definition}
Note that if instead $x_{\ell,i}>0$ in this definition,
then $x_{\ell,i}$ can be reduced and 
$x_{m,i}$  increased to reduce 
$ U_\ell -U_m$.
Also, if $\{x_{n,i}\}$ is $U$-MMF and  $x_{m,i},x_{\ell,i}>0$ 
for some server $i$ then
$U_m=U_\ell$.
Quantization (containerization)
issues associated with workload resource demands
are considered in \cite{Friedman14}.

Under multi-server DRF \cite{DRF,BLi15},
frameworks $n$ are selected using criterion
\be \label{global-DRF}
M_n 
& = & \frac{1}{\phi_n} 
x_n
\max_r \frac{d_{n,r}}{\sum_j c_{j,r}},
\ee
where $x_n = \sum_i x_{n,i}$.
That is, under multi-server DRF,
\be\label{DRF-u}
\forall i,~ u_{n,i} & = & \max_r \frac{d_{n,r}}{ \sum_j c_{j,r}}.
\ee

The server-specific PS-DSF criterion can be written as
\be \label{general-K}
K_{n,j} &=& \frac{\sum_i x_{n,i} d_{n,\rho(n,j)}}{\phi_n c_{j,\rho(n,j)}}
  ~=~  \frac{B_{n,j,\rho(n,j)}x_n}{\phi_n},
\ee 
where $\rho$ is such that
\be
B_{n,j,\rho(n,j)} & := & \max_r B_{n,j,r} ~~\mbox{when}~ \delta_{n,j}=1.
\label{rho-def}
\ee
Max-min fairness according to the joint framework-server
criterion $K_{n,j}$ is considered in 
\cite{PS-DSF-arxiv,PS-DSF-arxiv2,Jalal18}.
Here define
\be
K_n & = & \sum_i K_{n,i}\delta_{n,i} ~=~ 
\frac{1}{\phi_n} x_n \sum_i B_{n,i,\rho(n,i)}\delta_{n,i} \nonumber\\
&  = &\frac{1}{\phi_n} x_n \sum_i \max_r \frac{d_{n,r}}{c_{i,r}}\delta_{n,i}
\label{general-K2}
\ee
So, under PS-DSF,
\be\label{PS-DSF-u}
\forall i\in N_i,~ u_{n,i} 
&  = &\max_r \frac{d_{n,r}}{c_{i,r}}.
\ee

~\\
\begin{proposition}\label{claim:linear-mmf}
A solution $x=\{x_{n,i}\}$ of the optimization (\ref{Omega2}) s.t.
(\ref{no-overbook2})
has at least one resource $r$ fully booked in each server $i$.
In addition, there is a unique $U$-MMF solution 
if also:
\be \label{B-constraint}
\exists j ~\mbox{s.t.}~ \delta_{m,j}=1=\delta_{\ell,j}  & \Rightarrow &
\forall r,~ d_{m,r} = d_{\ell,r}, ~N_m=N_\ell, ~\mbox{and}~
\forall i, ~u_{m,i}=u_{\ell,i}.
\ee
\end{proposition}

{\em Proof:}  See Appendix A. The  proof is an adaptation of
that in  \cite{BG92,Ashari16b} for a single resource type.

~\\
Considering (\ref{DRF-u}) and (\ref{PS-DSF-u}),
$\forall r,~ d_{m,r} = d_{\ell,r}$ implies
$\forall i, u_{m,i}=u_{\ell,i}$.
So, (\ref{B-constraint}) is satisfied for both
DRF and PS-DSF when only frameworks with the same
resource demands share the same set of servers.


For task-based allocations (integer-valued $x$),
max-min fairness can be approximated by a greedy incremental
optimization known as progressive filling,
see \cite{BG92,DRF,Spark-Mesos-arxiv}.

\section{Proportional Fairness} \label{sec:PF}

For weighted proportional fairness, consider the objective
\be\label{Omega2b}
\max_{x}\sum_n \phi_n g_a(x_n),
\ee
\ie without dividing by $\phi_n$ in the argument of $g_a$ \cite{Mo00}.
For parameter $a>0$ specifically take 
\beqa
g_a(X) & = & \left\{\begin{array}{ll}
\log(X) & \mbox{if}~a=1\\
(1-a)^{-1}X^{1-a} & \mbox{else}
\end{array}\right.
\eeqa
\ie $g_a'(X) = 1/X^a$, again see \cite{Mo00}.
Obviously, in the case of $a=1$ ($g=\log$), whether
the factor $\phi$ is in the argument of $g$ is
immaterial.

The following generalizes Lemma 2 of \cite{Mo00} on Proportional Fairness.
See also the proportional-fairness/efficiency trade-off
framework of \cite{Chiang12} for a single server.

~\\
\begin{proposition}\label{claim:linear-prop}
A solution $x^*$ of the optimization (\ref{Omega2b}) s.t.
(\ref{no-overbook2}) is uniquely (weighted) 
$(\phi,a)$ $x$-proportional fair,
\ie for any other feasible solution $x$, 
\be\label{prop-efficient}
\Phi(x,x^*):=\sum_n \phi_n \frac{x_n - x_n^*}{(x_n^*)^a} & \leq & 0.
\ee
\end{proposition}

{\em Proof:}  See Appendix B.

~\\
From the proof, $\{x^*_n=\sum_i x^*_{n,i}\}_n$ is unique though
$x^*=\{x_{n,i}^*\}_{n,i}$ may not be. 
We can normalize $\hat{\phi}_n := \phi_n/\sum_k \phi_k$
and when $a=1$ write (\ref{prop-efficient}) as
\beqa
\sum_n \hat{\phi}_n \frac{x_n}{x_n^*} & \leq & 1.
\eeqa
A possible definition of the {\em efficiency}  of a
feasible allocation is  (\ref{revenue}) corresponding to
$a=0$, 
\be\label{eta-efficiency}
\sum_n \phi_n \sum_i x_{n,i} 
&=&\sum_n \phi_n x_n, 
\ee
\ie the
weighted total number of tasks scheduled. 
So, the optimization of 
Proposition \ref{claim:linear-prop} with $a=1$ gives an allocation
$x^*$ that is related to a task efficient allocation.
Clearly, $x^*$ satisfying
(\ref{prop-efficient}) for all other allocations $x$ with $a=1$
does not necessarily
maximize (\ref{eta-efficiency}). This issue is analogous to estimating
the mean of the ratio of positive
random variables $\E (X/X^*)$ using the ratio of the 
means $\E X/\E X^*$, see {\em e.g.} 
p. 351 of \cite{Kendall98} or
(11) of \cite{Seltman} .
For simplicity in the following, we use (\ref{eta-efficiency}) instead of
(\ref{prop-efficient}).

Note that the priority $\phi_n$
of framework $n$ could factor its resource footprint $\{d_{n,r}\}_r$.
Alternatively,
the resource footprints of the  frameworks
can be explicitly incorporated
into the main optimization objective via
a fairness criterion.
The proof of the following corollary is just as
that of Proposition \ref{claim:linear-prop}.
Recall that the generic fairness criterion $U_n$ (\ref{generic-criterion})
is a linear combination of $\{x_{n,i}\}_i$. 

~\\
\begin{corollary}\label{cor:U-prop}
A solution $x^*$ of the optimization problem
\beqa
\max_{x}\sum_n \phi_n \log(U_n) & \mbox{s.t.} &
(\ref{no-overbook2})
\eeqa
is uniquely $(\phi,1)$ $U$-proportional fair, \ie for any other feasible $x$,
\beqa
\sum_n \phi_n \frac{U_n - U_n^*}{U_n^*} & \leq & 0.
\eeqa
\end{corollary}

~\\
Again, optimal $\{U^*_n\}$ would be unique but
$x^*=\{x_{n,i}^*\}_{n,i}$ may not be.

Recall for DRF and PS-DSF,  the  
 $K_n$ (\ref{general-K2}) and 
 $M_n$ (\ref{global-DRF}), respectively,
are proportional to $x_n$.
Thus, using $U_n=K_n$ or $U_n=M_n$ in Corollary \ref{cor:U-prop} 
reduces to the result of Proposition \ref{claim:linear-prop}
when $a=1$.

\section{Progressive filling to 
approximate max-min fair allocation}\label{sec:progfill}

In the following evaluation studies, 
resources are incrementally (taskwise) allocated to 
frameworks $n$ with the intention to approximate  max-min fairness
(with respect to the fairness criterion used).
The approach is greedy: simply, the framework $n$ with 
smallest fairness criterion $U_n$ (or $U_{n,i}$), based on {\em existing}
allocations $\{x_{n,i}\}_{n,i}$, will be allocated a resource increment
$ \{\varepsilon d_{n,i}\}_i$ for small\footnote{Typically $\varepsilon=1$
when allocations $x$ are measured in ``tasks".}
$\varepsilon>0$.  
If a framework's resource demands
cannot be accommodated with available resources,
the framework with the next smallest fairness 
criterion will be
allocated  by this {\em progressive filling} approach
\cite{BG92,DRF}.
The  choice of server from which to allocate can be random,
\eg as for the Mesos default task-level progressive filling 
for DRF, 
see \cite{Spark-Mesos-arxiv}.
Alternatively, the framework and server
can be jointly chosen (\eg using PS-DSF).

Note how progressive filling can operate in the presence of
churn in the set of active frameworks, where in asynchronous fashion,
new frameworks could be initiated  or
a framework would
release all of its resources once its computations are completed, 
see \cite{Spark-Mesos-arxiv}.
In the following we assess the efficiencies of max-min fair approximations
by progressive filling according to different schedulers.

Because there is no resource revocation, a problem occurs when, say,
servers are booked so that 
there are insufficient spare resources to allocate
for a task of a just initiated framework (particularly a higher priority one).
Thus, new frameworks may need to wait for sufficient resources to be released
(by the termination of other frameworks).
Alternatively, {\em all} existing frameworks could be reallocated whenever 
any new framework initiates or any existing framework terminates.
Though within a server such reallocations are commonplace in a private setting,
the effect of such ``live" reallocations may be that tasks need to be terminated
and reassigned to other servers (or live migrated).
The following illustrative numerical examples allocate a single 
initial framework batch (without framework churn).
In the following emulation study for equal priority workloads and framework
churn, we 
work with the default progressive-filling mechanism  in Mesos wherein
existing frameworks are not adjusted upon framework churn.

\section{Evaluation objectives: Task efficiency of max-min fair allocations}\label{sec:evalobj}

In the following, though we aim for max-min fairness with progressive filling,
we are
also interested in the proportional fairness achieved.
We compare the efficiency (\ref{eta-efficiency})
of the allocations achieved by progressive filling
for examples with heterogeneous workloads and servers.
In the performance evaluation of our Mesos implementations,
efficiency is defined by overall execution time.

Though PS-DSF allocations achieved by progressive filling
may not be Pareto optimal, 
we show that they are more efficient,
even in some of our Mesos experiments where servers are
(at least initially) selected at random.

In the following, for brevity,
we consider only cases with frameworks of
equal priority 
($\forall n,n',~\phi_n=\phi_{n'}$)
and
without server-preference constraints
(\ie $\delta_{n,i}\equiv 1$).

\section{Illustrative numerical study of 
fair scheduling by progressive filling}\label{sec:numer}

In this section, we 
consider the following 
typical example of our numerical study
with two heterogeneous distributed application frameworks ($n=1,2$)
having resource demands per unit workload:
\be
d_{1,1} = 5,~ d_{1,2}=1,~ d_{2,1}=1,~ d_{2,2}=5; \label{illus-d}
\ee
and two heterogeneous servers ($i=1,2$) having two different
resources with capacities:
\be
c_{1,1}=100,~ c_{1,2}=30,~ c_{2,1}=30,~ c_{2,2}=100. \label{illus-c}
\ee 
For DRF and TSF, the servers $i$
are chosen in round-robin fashion,  where the server order is  randomly
permuted in each round; DRF under such randomized round-robin (RRR) 
server selection is
the default Mesos scheduler, \cf next section. 
One can also formulate PS-DSF under RRR wherein RRR selects the server
and the PS-DSF criterion only selects the framework for that server.
Frameworks $n$ are chosen by
progressive filling with integer-valued tasking ($x$), \ie whole
tasks are scheduled.

Numerical results for scheduled workloads
for this illustrative example are given in 
Tables \ref{table:illus} \& \ref{table:illus_stddev}, and
unused resources are given in 
Tables \ref{table:unused} and \ref{table:unused_stddev}.
200 trials were performed for DRF, TSF and PS-DSF under RRR
server selection, so using
Table \ref{table:illus_stddev} we can obtain confidence intervals
for the averaged quantities given in Table \ref{table:illus}
for schedulers under RRR. For example,  the 95\% confidence interval
for task allocation of the first framework on the second server 
(\ie $(n,i)=(1,2)$) 
under TSF is 
$$(6.5 -2 \cdot 0.46/\sqrt{200},6.5 +2\cdot 0.46/\sqrt{200}) 
= (6.43,6.57).$$
Note how PS-DSF's performance under RRR is comparable to 
when frameworks and servers are jointly selected \cite{Jalal18},
and with low variance in allocations.
We also found that RRR-rPS-DSF performed just as
rPS-DSF over 200 trials.

\begin{table}
\begin{center}
\begin{tabular}{|c||c|c|c|c||c|} 
\hline
\backslashbox{sched.}{$(n,i)$} &(1,1)&(1,2)&(2,1)&(2,2) & total \\ \hline\hline
DRF \cite{DRF,BLi15} & 6.55 & 4.69 & 4.69 & 6.55  & 22.48\\ \hline
TSF \cite{BLi16b} &  6.5 & 4.7  & 4.7 & 6.5 & 22.4\\ \hline
RRR-PS-DSF  & 19.44 & 1.15 & 1.07 & 19.42 & 41.08 \\ \hline 
BF-DRF \cite{BLi15} & 20 & 2 & 0 & 19 & 41\\ \hline
PS-DSF \cite{Jalal18} & 19 & 0 & 2 & 20 & 41 \\ \hline
rPS-DSF  & 19 & 2 & 2 & 19 & 42 \\ \hline 
\end{tabular}
\caption{Workload allocations $x_{n,i}$ for different schedulers
under progressive filling for illustrative
example with parameters (\ref{illus-d}) and
(\ref{illus-c}). Averaged values over 200 trials reported for
the first three schedulers operating under RRR
server selection.}\label{table:illus}
\end{center}
\end{table}

\begin{table}
\begin{center}
\begin{tabular}{|c||c|c|c|c|} 
\hline
\backslashbox{sched.}{$(n,i)$} &(1,1)&(1,2)&(2,1)&(2,2)   \\ \hline\hline
DRF \cite{DRF,BLi15} & 2.31 & 0.46 & 0.46 & 2.31 \\ \hline
TSF \cite{BLi16b} &  2.29 & 0.46 &  0.46 & 2.29 \\ \hline
RRR-PS-DSF        & 0.59 & 0.99 & 1 & 0.49   \\ \hline 
\end{tabular}
\caption{Sample standard deviation 
of allocations $x_{n,i}$ for different schedulers
under RRR server selection with. Averaged  values over 200
trials reported.}\label{table:illus_stddev}
\end{center}
\end{table}

\begin{table}
\begin{center}
\begin{tabular}{|c||c|c|c|c|} 
\hline
\backslashbox{sched.}{$(i,r)$} &(1,1)&(1,2)&(2,1)&(2,2)  \\ \hline\hline
DRF \cite{BLi15} & 62.56  & 0 & 0  &62.56  \\ \hline
TSF \cite{BLi16b} &  62.8 & 0 &  0 & 62.8 \\ \hline
RRR-PS-DSF & 1.8 & 4.6 & 4.86 & 1.92  \\ \hline 
BF-DRF \cite{BLi15} & 0 & 10 & 1 &  3 \\ \hline
PS-DSF \cite{Jalal18} & 3 & 1  & 10 & 0 \\ \hline
rPS-DSF  & 3 & 1 & 1 & 3 \\ \hline 
\end{tabular}
\caption{Unused capacities
$c_{i,r}-\sum_n x_{n,i}d_{i,r}$ for different schedulers
under progressive filling for illustrative
example with parameters (\ref{illus-d}) and
(\ref{illus-c}). Averaged values over 200 trials reported
under RRR server selection.}\label{table:unused}
\end{center}
\end{table}

\begin{table}
\begin{center}
\begin{tabular}{|c||c|c|c|c|} 
\hline
\backslashbox{sched.}{$(i,r)$} &(1,1)&(1,2)&(2,1)&(2,2)  \\ \hline\hline
DRF \cite{DRF,BLi15} & 11.09& 0 & 0  & 11.09   \\ \hline
TSF \cite{BLi16b} & 10.99  & 0 & 0   & 10.99  \\ \hline
RRR-PS-DSF        & 0.59 & 0.99 & 1 & 0.49  \\ \hline 
\end{tabular}
\caption{Sample standard deviation 
of unused capacities $c_{i,r}-\sum_n x_{n,i}d_{i,r}$ for different schedulers
under RRR server selection over 200 trials.}\label{table:unused_stddev}
\end{center}
\end{table}

We found task efficiencies improve using
{\em residual} forms of the fairness criterion.
For example, the residual  PS-DSF (rPS-DSF) criterion is
\beqa
\tK_{n,j,\ux_j} & = & x_n \max_r \frac{d_{n,r}}
{\phi_n( c_{j,r}-\sum_{n'} x_{n',j}d_{n',r})}
\eeqa
That is, this criterion makes scheduling decisions by 
progressive filling using {\em current  residual}
(unreserved) capacities based on the {\em current} allocations $x$.
From Table \ref{table:illus}, we see the improvement is
modest for the case of PS-DSF.

Improvements are also obtained by {\em best-fit} 
server selection. For example, best-fit DRF  (BF-DRF) 
first selects framework $n$ by DRF and then 
selects the server whose residual
capacity most closely matches their
resource demands $\{d_{n,r}\}_r$ \cite{BLi15}.

\section{Summary and Future Work}\label{sec:summary}
For a private-cloud setting, we considered scheduling
a group of heterogeneous, distributed frameworks
to a group of heterogeneous servers.
We extended two general results on max-min fairness and
proportional fairness to this case for a static
problem under generic scheduling criteria.
Subsequently, we assessed the efficiency of approximate
max-min fair allocations by progressive filling according
to different fairness criteria. 
Illustrative examples in heterogeneous settings  show that
max-min fair PS-DSF and rPS-DSF scheduling,
are superior to DRF in terms of 
task efficiency performance (a metric related to
proportional fairness) and that the efficiency
of these  ``server  specific" schedulers 
did not significantly suffer from the use of
randomized round-robin server selection.
Task efficiency was also
improved  when either the ``best fit" approach to selecting
servers was used or the fairness criteria was modified
to use current residual/unreserved resource capacities.
We also open-source implemented
oblivious (``coarse grained") and workload-characterized
(specified resource demands $d$) {\em online}
prototypes of these schedulers on 
Mesos \cite{mesos-code,spark-code},
with the Mesos default/baseline being oblivious DRF. 
Using two different Spark workloads and heterogeneous
servers, we showed that the schedulers were
similarly ranked using the total execution time as
the performance measure.  Moreover,  execution times
could be shortened with workload characterization.

In future work, we will consider scheduling 
(admission control and placement) problems in a 
public cloud setting. To this end,
note that similar objectives to those considered
herein for a private-cloud setting, particularly (\ref{revenue}), 
may be reinterpreted as overall revenue
based on {\em bids} $\phi$ for virtual machines or containers with 
fixed resource allocations  $d$.
Also, as profit margins diminish in a maturing marketplace,
one expects that public clouds will need to operate
with greater resource efficiency.
Note that notions of fair scheduling
and desirable properties of schedulers as defined in, 
\eg \cite{DRF,Friedman14,BLi16} may not
be relevant to the public-cloud setting, where the expectation is
that different customers/frameworks simply ``get what they pay for." 
Moreover, in a public cloud setting, what the customers
do with their virtual machines/containers is arguably not the concern
of the cloud operator so long as the customer complies with service level
agreements.   But, \eg notions of strategy proofness
are important considerations in the design of
auction \cite{VCG-auction} and spot-pricing mechanisms
(where under spot price mechanisms, virtual machines or containers
may be revoked).

\bibliographystyle{abbrv}
\bibliography{../../latex/scheduling,../../latex/cloud,../../latex/opt,../../latex/mars,../../latex/refs-cloud,../../latex/kesidis-prior,../../latex/IPM,../../latex/ref_cheng,../../latex/cloud2,../../latex/knapsack,../../latex/games,../../latex/ratio}

\begin{thebibliography}{10}

\bibitem{Abdelzaher:2002:PGW:506156.506162}
T.~F. Abdelzaher, K.~G. Shin, and N.~Bhatti.
\newblock Performance guarantees for web server end-systems: A
  control-theoretical approach.
\newblock {\em IEEE Trans. Parallel Distrib. Syst.}, 13(1):80--96, 2002.

\bibitem{BG92}
D.~Bertsekas and R.~Gallager.
\newblock {\em Data Networks, 2nd Ed.}
\newblock Prentice Hall, 1992.

\bibitem{Chandra:2003:DRA:781027.781067}
A.~Chandra, W.~Gong, and P.~Shenoy.
\newblock Dynamic resource allocation for shared data centers using online
  measurements.
\newblock In {\em Proceedings of the 2003 ACM SIGMETRICS International
  Conference on Measurement and Modeling of Computer Systems}, SIGMETRICS '03,
  2003.

\bibitem{CK04}
C.~Chekuri and S.~Khanna.
\newblock On multi-dimensional packing problems.
\newblock {\em SIAM Journal of Computing}, 33(4):837--851, 2004.

\bibitem{HUG}
M.~Chowdhury, Z.~Liu, A.~Ghodsi, and I.~Stoica.
\newblock {HUG: Multi-resource fairness for correlated and elastic demands}.
\newblock In {\em Proc. USENIX NSDI}, March 2016.

\bibitem{CKPT16}
H.~Christensen, A.~Khan, S.~Pokutta, and P.~Tetali.
\newblock {Multidimensional Bin Packing and Other Related Problems: A Survey}.
\newblock https://people.math.gatech.edu/$\sim$tetali/PUBLIS/CKPT.pdf, 2016.

\bibitem{Cohen:2004:CID:1251254.1251270}
I.~Cohen, M.~Goldszmidt, T.~Kelly, J.~Symons, and J.~S. Chase.
\newblock Correlating instrumentation data to system states: A building block
  for automated diagnosis and control.
\newblock In {\em Proceedings of the 6th Conference on Symposium on Opearting
  Systems Design \& Implementation - Volume 6}, OSDI'04, 2004.

\bibitem{Cohen17}
M.~Cohen, V.Mirrokni, P.~Keller, and M.~Zadimoghaddam.
\newblock {Overcommitment in Cloud Services Bin packing with Chance
  Constraints}.
\newblock In {\em Proc. ACM SIGMETRICS}, Urbana-Campaign, IL, June 2017.

\bibitem{Doyle:2003:MRP:1251460.1251465}
R.~P. Doyle, J.~S. Chase, O.~M. Asad, W.~Jin, and A.~M. Vahdat.
\newblock Model-based resource provisioning in a web service utility.
\newblock In {\em Proceedings of the 4th Conference on USENIX Symposium on
  Internet Technologies and Systems - Volume 4}, USITS'03, 2003.

\bibitem{duke}
{Duke utility bill tariff}, 2012.
\newblock \url{http://www.considerthecarolinas.com/pdfs/scscheduleopt.pdf}.

\bibitem{Friedman14}
E.~Friedman, A.~Ghodsi, and C.-A. Psomas.
\newblock Strategyproof allocation of discrete jobs on multiple machines.
\newblock In {\em Proc. ACM Conf. on Economics and Computation}, 2014.

\bibitem{DRF}
A.~Ghodsi, M.~Zaharia, B.~Hindman, A.~Konwinski, S.~Shenker, and I.~Stoica.
\newblock Dominant resource fairness: Fair allocation of multiple resource
  types.
\newblock In {\em Proc. USENIX NSDI}, 2011.

\bibitem{Mesos}
B.~Hindman, A.~Konwinski, M.~Zaharia, A.~Ghodsi, A.~Joseph, R.~Katz,
  S.~Shenker, and I.~Stoica.
\newblock {Mesos: A Platform for Fine-grained Resource Sharing in the Data
  Center}.
\newblock In {\em Proc. USENIX NSDI}, 2011.

\bibitem{Chiang12}
C.~Joe-Wong, S.~Sen, T.~Lan, and M.~Chiang.
\newblock {Multi-resource allocation: Fairness-efficiency tradeoffs in a
  unifying framework}.
\newblock {\em IEEE/ACM Trans. Networking}, 21(6), Dec. 2013.

\bibitem{Ashari16b}
J.~Khamse-Ashari, G.~Kesidis, I.~Lambadaris, B.~Urgaonkar, and Y.~Zhao.
\newblock {Constrained Max-Min Fair Scheduling of Variable-Length Packet-Flows
  to Multiple Servers}.
\newblock In {\em Proc. IEEE GLOBECOM}, Washington, DC, Dec. 2016.

\bibitem{PS-DSF-arxiv2}
J.~Khamse-Ashari, I.~Lambadaris, G.~Kesidis, B.~Urgaonkar, and Y.~Zhao.
\newblock {An Efficient and Fair Multi-Resource Allocation Mechanism for
  Heterogeneous Servers}.
\newblock http://arxiv.org/abs/1712.10114, Dec. 2017.

\bibitem{Jalal18}
J.~Khamse-Ashari, I.~Lambadaris, G.~Kesidis, B.~Urgaonkar, and Y.~Zhao.
\newblock {An Efficient and Fair Multi-Resource Allocation Mechanism for
  Heterogeneous Servers}.
\newblock {\em IEEE Trans. Parallel and Distributed Systems (TPDS)}, May 2018.

\bibitem{PS-DSF-arxiv}
J.~Khamse-Ashari, I.~Lambadaris, G.~Kesidis, B.~Urgaonkar, and Y.~Zhao.
\newblock {Per-Server Dominant-Share Fairness (PS-DSF): A Multi-Resource Fair
  Allocation Mechanism for Heterogeneous Servers}.
\newblock https://arxiv.org/abs/1611.00404, Nov. 1, 2016.

\bibitem{Levy2003}
R.~Levy, J.~Nagarajarao, G.~Pacifici, M.~Spreitzer, A.~Tantawi, and A.~Youssef.
\newblock Performance management for cluster based web services.
\newblock In G.~Goldszmidt and J.~Sch{\"o}nw{\"a}lder, editors, {\em Integrated
  Network Management VIII: Managing It All}, pages 247--261. Springer US, 2003.

\bibitem{Lu:2001:FCA:882481.883781}
C.~Lu, T.~F. Abdelzaher, J.~A. Stankovic, and S.~H. Son.
\newblock A feedback control approach for guaranteeing relative delays in web
  servers.
\newblock In {\em Proceedings of the Seventh Real-Time Technology and
  Applications Symposium}, RTAS '01, 2001.

\bibitem{Menasce:2003:WSS:1050672.1050719}
D.~A. Menasce.
\newblock Web server software architectures.
\newblock {\em IEEE Internet Computing}, 7(6):78--81, 2003.

\bibitem{mesos-code}
Mesos multi-scheduler.
\newblock https://github.com/PSU-Cloud/mesos-ps/pull/1/files.

\bibitem{Mo00}
J.~Mo and J.~Walrand.
\newblock Fair end-to-end window-based congestion control.
\newblock {\em IEEE/ACM Trans. Networking}, Vol. 8, No. 5:pp. 556--567, 2000.

\bibitem{N.Bennani:2005:RAA:1078027.1078472}
M.~N.~Bennani and D.~A.~Menasce.
\newblock Resource allocation for autonomic data centers using analytic
  performance models.
\newblock In {\em Proceedings of the Second International Conference on
  Automatic Computing}, ICAC '05. IEEE Computer Society, 2005.

\bibitem{Seltman}
H.~Seltman.
\newblock Approximation of mean and variance of a ratio.
\newblock http://www.stat.cmu.edu/~hseltman/files/ratio.pdf.

\bibitem{Spark-Mesos-arxiv}
Y.~Shan, A.~Jain, G.~Kesidis, B.~Urgaonkar, J.~Khamse-Ashari, and
  I.~Lambadaris.
\newblock {Online Scheduling of Spark Workloads with Mesos using Different Fair
  Allocation Algorithms}.
\newblock https://arxiv.org/abs/1803.00922, March 2, 2018.

\bibitem{spark-code}
{Spark with HeMT}.
\newblock https://github.com/PSU-Cloud/spark-hemt/pull/2/files.

\bibitem{Kendall98}
A.~Stuart and K.~Ord.
\newblock {\em Kendall's Advanced Theory of Statistics}.
\newblock Arnold, London, 6th edition, 1998.

\bibitem{VCG-auction}
{Vickrey-Clarke-Groves auction}.
\newblock https://en.wikipedia.org/wiki/Vickrey-Clarke-Groves\_auction.

\bibitem{BLi16}
W.~Wang, B.~Li, B.~Liang, and J.~Li.
\newblock Towards multi-resource fair allocation with placement constraints.
\newblock In {\em Proc. ACM SIGMETRICS}, Antibes, France, 2015.

\bibitem{BLi16b}
W.~Wang, B.~Li, B.~Liang, and J.~Li.
\newblock Multi-resource fair sharing for datacenter jobs with placement
  constraints.
\newblock In {\em Proc. Supercomputing}, Salt Lake City, Utah, 2016.

\bibitem{BLi15}
W.~Wang, B.~Liang, and B.~Li.
\newblock Multi-resource fair allocation in heterogeneous cloud computing
  systems.
\newblock {\em IEEE Transactions on Parallel and Distributed Systems},
  26(10):2822--2835, Oct. 2015.

\bibitem{159}
W.~Xu, P.~Bodik, and D.~Patterson.
\newblock A flexible architecture for statistical learning and data mining from
  system log streams.
\newblock In {\em Proceedings of Workshop on Temporal Data Mining: Algorithms,
  Theory and Applications at the Fourth IEEE International Conference on Data
  Mining}, Brighton, UK, 2004.

\bibitem{McKeown13}
K.-K. Yap, T.-Y. Huang, Y.~Yiakoumis, S.~Chinchali, N.~McKeown, and S.~Katti.
\newblock Scheduling packets over multiple interfaces while respecting user
  preferences.
\newblock In {\em Proc. ACM CoNEXT}, Dec. 2013.

\end{thebibliography}

\section*{Appendix A: Proof of Proposition \ref{claim:linear-mmf}}

Define the Lagrangian  to be maximized over $x$
and over Lagrange multipliers $\lambda,\nu \geq 0$:
\beqa
L & = & \sum_n \phi_n g(U_n)
+ \sum_{i,r}\lambda_{i,r}(1-\sum_{n\in N_i}x_{n,i}B_{n,i,r})\\
& & ~~
+ \sum_{i,n\in N_i}\nu_{n,i}x_{n,i}.
\eeqa
The first-order optimality condition,
\be
\lefteqn{\forall i,n\in N_i,~\delta_{n,i}=1,} & & \nonumber\\
0   =  \frac{\partial L}{\partial x_{n,i}}
& = & u_{n,i}g'(U_n) - \sum_r \lambda_{i,r}B_{n,i,r}+\nu_{n,i},
\label{FOOC2}
\ee
and $g$ strictly increasing 
imply 
\be\label{lambda-nu-bound}
\forall i,n\in N_i, ~
 \sum_r \lambda_{i,r}B_{n,i,r}> \nu_{n,i}\geq 0.
\ee
So, $\forall i$, 
$\exists r$ s.t. $\lambda_{i,r}> 0$.
Thus, complementary slackness is
\be 
\forall i,r,~ \lambda_{i,r}(1-\sum_{n\in N_i}x_{n,i}B_{n,i,r})& = &0 
\label{CS1}\\
\Rightarrow~~ \forall i, \exists r ~\mbox{s.t.} ~  \sum_{n\in N_i}x_{n,i} B_{n,i,r} & = &1,
\label{fully-booked}
\ee
\ie  in every server $i$, one resource $r$ (which may depend on $i$)
is fully booked.  
So, the set of fully
booked resources in server $i$ under allocations
$x=\{x_{n,i}\}$ can be characterized by
$\{r ~|~\lambda_{i,r}>0\}$.
Now by (\ref{FOOC2}) and assumed strict concavity of $g$,   uniquely
\beqa
\forall i, n\in N_i,~
U_n & = & 
(g')^{-1}\left(\sum_r \lambda_{i,r}\frac{B_{n,i,r}}{u_{n,i}}-
\frac{\nu_{n,i}}{u_{n,i}}\right)\\
& = & 
(g')^{-1}\left(\sum_{r:\lambda_{i,r}>0} \lambda_{i,r}\frac{B_{n,i,r}}{u_{n,i}}-
\frac{\nu_{n,i}}{u_{n,i}}\right).
\eeqa
Now consider two frameworks $m$ and $\ell$ and server
$i$ such that  $x_{m,i}>0$
and $\delta_{m,i}=1=\delta_{\ell,i}$. 
So, complementary slackness  
\be\label{CS2}
\forall j, n\in N_j,~
 \nu_{n,j}x_{n,j} & = & 0,
\ee
implies $\nu_{m,i}=0$.

Because $(g')^{-1}$ is strictly decreasing 
($g$ strictly concave): if $\delta_{m,i}=1=\delta_{\ell,i}$ then
\beqa
U_m& = & (g')^{-1}\left(\sum_{r:\lambda_{i,r}>0}
 \lambda_{i,r}\frac{B_{m,i,r}}{u_{m,i}}
\right)\\
&\leq & 
(g')^{-1}\left(\sum_{r: \lambda_{i,r}>0} \lambda_{i,r}
\frac{B_{\ell,i,r}}{u_{\ell,i}}-\frac{\nu_{\ell,i}}{u_{\ell,i}}\right) ~=~ 
U_\ell,
\eeqa
where we have used assumption (\ref{B-constraint}) which is
sufficient  for the inequality.
Because of this and (\ref{fully-booked}), 
a solution $x=\{x_{n,i}\}$ of the optimization (\ref{Omega2}) s.t.
(\ref{no-overbook2}) is $U$-MMF.

\section*{Appendix B: Proof of  Proposition \ref{claim:linear-prop}}

The Lagrangian here is
\beqa
L & = & \sum_n \phi_n g_a(x_n)  
+ \sum_{i,r}\lambda_{i,r}(1-\sum_{n\in N_i}x_{n,i}B_{n,i,r})\\
& & ~~ + \sum_{i, n\in N_i}\nu_{n,i}x_{n,i}\delta_{n,i}
\eeqa
where, again, the Lagrange multipliers  $\lambda,\nu\geq 0$.
A first-order optimality condition is
\be
\forall i, n\in N_i,~
0  & = &  \frac{\partial L}{\partial x_{n,i}} (x^*) \nonumber\\
 & = & \phi_n g_a'(x_n^*) - \sum_r \lambda_{i,r}B_{n,i,r}+\nu_{n,i}.
\label{FOOC2b}
\ee

Multiplying  (\ref{FOOC2b}) by 
$x_{n,i}-x_{n,i}^*$  
and summing over $i$ and $n\in N_i$ gives\footnote{Simply 
use Fubini's theorem for the first term,
$\sum_i \sum_{n\in N_i} \phi_n g'(x_n^*)(x_{n,i}-x_{n,i}^*) = 
\sum_n \sum_{i:\delta_{n,i}=1} \phi_n g'(x_n^*)(x_{n,i}-x_{n,i}^*)$}
\beqa
0 & = & \sum_n \phi_n g_a'(x_n^*)(x_n-x_n^*)  
+\sum_{i,n\in N_i} \nu_{n,i}(x_{n,i}-x_{n,i}^*)\\
& & ~~-\sum_{i,r} \sum_{n\in N_i}
\lambda_{i,r}B_{n,i,r}(x_{n,i}-x_{n,i}^*)
\eeqa
where the first term is $\Phi(x,x^*)$
and recall the definition of $N_i$ (\ref{x-sum-def}).
Thus, by complementary slackness (\ref{CS1}) and (\ref{CS2}) (taking
$x=x^*$ there in those equations),
\beqa
\Phi(x,x^*)  = 
\sum_{i,r:\lambda_{i,r}\not=0}\lambda_{i,r}\left(\sum_{n\in N_i} x_{n,i}B_{n,i,r}-1\right)
-\sum_{i,n\in N_i} \nu_{n,i}x_{n,i}.
\eeqa
Finally,  no resource overbooking
 (\ref{no-overbook2}) implies
$\Phi(x,x^*)\leq 0$.

\end{document}